\documentclass[preprintnumbers,aps,prd,floatfix,nofootinbib,onecolumn]{revtex4}
\usepackage{qcd}
\usepackage{epsfig}
\usepackage{graphicx}
\usepackage[normalem]{ulem}
\usepackage[pagebackref]{hyperref}
\usepackage{bm}
\usepackage{float}
\usepackage{orcidlink}
\usepackage{ifthen}
\usepackage{etoolbox}

\definecolor{nicegreen}{rgb}{0.,0.5,0.}

\newcommand{\vers}{arx}

\begin{document}

\title{Reply to ``Comment on 
`QCD factorization with multihadron fragmentation \\ functions' ''}

\author{T.~C.~Rogers \orcidlink{0000-0002-0762-0275}\,}
\email{trogers@odu.edu}
\affiliation{Department of Physics, Old Dominion University, Norfolk, VA 23529, USA}
%

%
\author{M.~Radici \orcidlink{0000-0002-4542-9797}\,}
\email{marco.radici@pv.infn.it}
\affiliation{INFN - Sezione di Pavia, via Bassi 6, I-27100 Pavia, Italy}

\author{A.~Courtoy\orcidlink{0000-0001-8906-2440}\,}
\email{aurore@fisica.unam.mx}
\affiliation{Instituto de F\'isica,
  Universidad Nacional Aut\'onoma de M\'exico, Apartado Postal 20-364,
  01000 Ciudad de M\'exico, Mexico}

\author{T.~Rainaldi \orcidlink{0000-0002-8342-6765}\,}
\email{tommaso.rainaldi@stonybrook.edu}
\affiliation{Department of Physics, Old Dominion University, Norfolk, VA 23529, USA}

\begin{abstract}
We respond to comments in Ref.~\cite{Pitonyak:2025lin} about our article, Ref.~\cite{Rogers:2024nhb}.   We stand by our conclusions and defend them against the criticisms.
\end{abstract}

\date{\today}

\maketitle

\section{Introduction}

 An important feature of factorization theorems like Ref.~\cite[Eq.~(3)]{Rogers:2024nhb} is that all dependence on the observed small-mass nonperturbative external states is isolated within the states themselves. When considering different external states, one simply switches $| h \rangle \to | h' \rangle$ in the relevant parton correlation function and nothing else in the formula changes. Our example, collinear factorization for the semi-inclusive annihilation (SIA) process, is among the most firmly established of all the factorization formulas. No new derivations are necessary when changing the identity of the observed final state, as long as the kinematics remain consistent with collinear SIA as described by a single fragmenting parton. 

Historically, dihadron fragmentation functions have been very attractive for studying nonperturbative structures because the above means the factorization theorem applies automatically to nonperturbative hadronic final states made of multiple hadrons, so long as one remains within the kinematical domain of factorization. 
 That is, it is valid for the $\diff{Y}{}$ in Ref.~\cite[Eq.~(3)]{Rogers:2024nhb} to be the general phase space in Ref.~\cite[Eq.~(4)]{Rogers:2024nhb}.  All results can be taken over directly from existing knowledge about the treatment of single hadron factorization. 
See, for example, the textbook account in Sec.~12.13.3 of Ref.~\cite{Collins:2011qcdbook}. There one reads 
 \begin{quote}
 ``\textit{In the final state in a definition like (12.35), }\textit{we replace} $|p,X,\text{out}\rangle\langle\text{out},X,p|$ \textit{by} \newline $|p_1,p_2,X,\text{out}\rangle\langle\text{out},X,p_1,p_2|$. \textit{At the partonic end of the fragmentation function, nothing changes. So all the issues about renormalization, DGLAP evolution, and the construction of a hard-scattering coefficient function are unchanged.}'' 
 \end{quote}
 We see this as central when the interpretation of measurements involves intrinsic partonic structures. It is why, in past work, single and multihadron fragmentation functions have always been taken to have identical definitions, up to the identity of the nonperturbative final state -- see for example Ref.~\cite{Collins:1994ax}. 

However, 
Ref.~\cite{Pitonyak:2023gjx}   
makes strong claims that call into question this overall view, arguing that the standard fragmentation function definition has no number density interpretation at all when extended to  multihadrons, thus bringing into question the reliability of past work.
Instead, it argues that the basic operator definition
must be modified with nontrivial external-state-dependent prefactors. This is the origin of the current debate.

We disagree with these conclusions, and in our Ref.~\cite{Rogers:2024nhb} we defended the standard form of the definition and its origins in factorization. 
We identified the contradiction between it and the assertions of Ref.~\cite{Pitonyak:2023gjx} as originating from a mixing up of partonic and hadronic kinematics combined with the imposition of an invalid sum rule. 
We also explained that the ordinary fragmentation function does indeed retain a solid number density interpretation. 
In their comment~\cite{Pitonyak:2025lin}, the authors further promote their arguments from Ref.~\cite{Pitonyak:2023gjx}. 

The foundation of the argument in Ref.~\cite{Pitonyak:2023gjx} is to replace the regular factorization theorem for hadron multiplicities by a specific type of conjectured multiplicity sum rule \cite[Eq.~(6)]{Pitonyak:2023gjx}. 
Then, by demanding that this sum rule be satisfied, it is argued that the normal definition of the fragmentation function must be modified by a nontrivial prefactor when the final state is a small-mass multihadron system.

However, the sum rule in Ref.~\cite[Eq.~(6)]{Pitonyak:2023gjx} is not valid for multiple reasons. These were explained in Refs.~\cite{Collins:2023cuo,Rogers:2024zvi} and Ref.~\cite[Section VIII]{Rogers:2024nhb}. Moreover, it implies for the SIA cross section a factorized expression that contradicts the above principle, as recalled from Ref.~\cite{Collins:2011qcdbook} (and stressed also in Ref.~\cite{Rogers:2024nhb}). 
Our perspective is that
the sum rule is not a valid basis for either derivations or their interpretations. It is the root of the discrepancy between the standard definition and the one recommended in Refs.~\cite{Pitonyak:2023gjx}. We will continue to expand on this here.

\section{Parton Model}

For notation and an explanation of reference frames, see Ref.~\cite[Sec.III]{Rogers:2024nhb}.
\par\noindent
In Ref.~\cite{Pitonyak:2025lin} it is written that in Ref.~\cite{Rogers:2024nhb} we must ``\textit{define a “reduced” DiFF \underline{in the middle of} a parton model calculation}'' to ensure ``\textit{the expected
expression for the cross section is achieved.}'' 
That this is not the case can be seen from~\cite[Eq.~(86)]{Rogers:2024nhb}. By definition, the parton model is the result of dropping $\order{\alpha_s}$ errors (and integrating over angles for the unpolarized case) to get
\begin{equation}
\label{e.generaldz}
\frac{1}{\sigma_0} \frac{\diff{\sigma}{}}{ \diff{z} \diff{Y'}} \stackrel{\text{parton model}}{=}  d(z,\left\{p_h\right\}) \, .
\end{equation}
The $d(z,\left\{p_h\right\})$ is just the standard fragmentation function definition for a multi-hadron system with small invariant mass, collective 4-momentum variables $\{ p_h\}$, and carrying a fraction $\xi = z$ of the fragmenting parton momentum. It is completely determined, up to a choice of renormalization scheme, by its operator definition in ~\cite[Eqs.~(79-81)]{Rogers:2024nhb}. There is no need to invoke a concept of a reduced fragmentation function to get to the parton model expression here. 
For the transverse-momentum dependent (TMD) case, each combination of $\zetasym$, $-\zetasym \T{k}{H}$, 
and set of 4-vectors $\left\{ p_h \right\}$ maps onto a definite value for the fragmentation function, and the collinear function is obtained by integrating upon $-\zetasym \T{k}{H}$. In the parton model, $d(z,\left\{p_h\right\})$ is a density in $\diff{z} \diff{Y'}{}$. 

$\diff{Y'}$ is simply 
the result of transforming variables in the full hadronic phase space
as (see Ref.~\cite[Eq.~(4)]{Rogers:2024nhb})
 \begin{equation}
\diff{Y} = \frac{1}{2 E_{p_h} (2 \pi)^3} \diff[3]{\vect{p}_h}{} \diff{Y'} \, .
 \end{equation}
We could leave \eref{generaldz} differential in $\diff{Y}$, but separating out the total momentum $\vect{p}_h$ is convenient and allows us to immediately integrate over $\vect{p}_h$ angles. 
If writing a generic $\diff{Y'}{}$ is unsatisfying, one may use the specific variables in ~\cite[Eq.~(106)]{Rogers:2024nhb}: $\diff{Y'}{} = \diff{M_h}{} \diff{\zeta}{} \diff{\phi_{R,H}} \, M_h / 32 \pi^3 $. 

The above expressions 
are unambiguous. 
The fragmentation function is not a density with respect to an unspecified variable choice. Rather, it is a density with respect to the physical hadronic phase space from asymptotic hadronic states, the same states that appear when defining an $S$-matrix.
In terms of partons, it is a function of one partonic momentum fraction $\xi$, as expected because there is only one fragmenting parton leaving the hard part. 
The hard part (which here is $\sigma_0$) is independent of the arbitrary choice of external kinematical variables that describe hadrons. If in \eref{generaldz} one instead used the fragmentation function from Refs.\cite{Pitonyak:2023gjx,Pitonyak:2025lin} with its hadron-number-dependent prefactor, then the hard part would
also become hadron-number-dependent, thus breaking factorization. We explain this around our \cite[Eq.~(137)]{Rogers:2024nhb}.

The number density interpretation of the standard $d(z,\left\{p_h\right\})$ follows immediately from \eref{generaldz}. We hope this definitively clarifies that the standard definition does not lack a number density interpretation. In particular, the statement from Ref.~\cite{Pitonyak:2023gjx}
\begin{quote}
\textit{If one instead were to use a prefactor of $1/(4z)$, to be in full analogy with single-hadron fragmentation [1, 59–61], the quark uDiFFs would not retain a
number density interpretation} 
\end{quote}
is not correct.

When making variable transformations to the left side of \cite[Eq.~(3)]{Rogers:2024nhb}, there is no reference to partonic momenta. So, in deciding on a set of variables to use for external physical momenta, Jacobian factors should not involve internal partonic degrees of freedom in any way. The parton momentum is an internal integration variable. 
Therefore, we do not agree that the various \emph{partonic} variable transformations in Ref.~\cite[Sec.~III ]{Pitonyak:2025lin} are necessary or helpful.

If the relative internal momenta in an $n$-hadron system cannot be measured (but are restricted to the factorization region) then it is useful to rewrite \eref{generaldz} as
\begin{equation}
\label{e.generaldzb}
\frac{1}{\sigma_0} \frac{\diff{\sigma}{}}{ \diff{z}} \stackrel{\text{parton model}}{=}  d(z,\mathcal{V}) \, ,
\end{equation}
with
\begin{equation}
\label{e.generaldzc}
d(z,\mathcal{V}) = \int_\mathcal{V} \diff{Y'} d(z,\{ p_h \}) \, ,
\end{equation}
where $\mathcal{V}$ is some fixed region of the internal $n$-hadron phase space.
The $d(z,\mathcal{V})$ and $d(z,\{ p_h \})$ are the same fragmentation function definitions except that with $d(z,\mathcal{V})$ the final states include integrals 
upon relative internal momenta over some fixed region of phase space.
Most importantly, no factors get shifted between the hard part and the fragmentation function just because a subset of the final states includes integrals upon $\diff{Y'}{}$. As always, changing the identity of the measured final state changes nothing in the definition but the state itself. 

Leaving subsets of the variables in $\diff{Y'}{}$ unintegrated gives versions of \eref{generaldzc} and \eref{generaldz} differential in 
these subsets.
For example, leaving the integral over $M_h$ in \eref{generaldzc} undone immediately makes \eref{generaldzb} into
Eq.~(9) from Ref.~\cite{Courtoy:2012ry}.

\section{Factorization}

In our~\cite[Eq.~(137)]{Rogers:2024nhb} and surrounding text, we observed that one may force the fragmentation function preferred by  Refs.~\cite{Pitonyak:2023gjx,Pitonyak:2025lin} into an expression that matches the cross section, but only if 
at least an additional extra power of $\zetasym$ is allowed to be absorbed into the hard part for each extra hadron that appears in the final state. We wrote,
\begin{quote}
\textit{ ...one is forced to modify the hard partonic
part with extra powers of $\xi$ for each additional hadron}
\end{quote}
But having an external-state-dependent hard part destroys factorization, and then
the question of which prefactor for $d(\zetasym,\left\{p_h\right\})$ is correct becomes ill-defined. It means one may simply shift external-state-dependent factors arbitrarily between the hard part and the fragmentation function 
without invalidating the expression for the cross section. 
That is, one may always break
factorization in the formula~\cite[Eq.(3)]{Rogers:2024nhb} just by rewriting it as
\begin{equation}
\label{e.fact_formulapf}
\frac{\diff{\sigma}{}}{\diff{Y}{}} = \int_z^1 \frac{\diff{\zetasym}{}}{\zetasym^2} \parz{X_h(\zetasym)  2 E_{\hat{k}} (2 \pi)^3 \frac{\diff{\hat{\sigma}}{}}{\diff[3]{\vect{\hat{k}}}{}} } \underbrace{\parz{\frac{1}{X_h(\zetasym)} d(\zetasym,\left\{p_h\right\})}}_{d'(\zetasym,\left\{p_h\right\})}  \, ,
\end{equation}
where $X_h(\zetasym)$ is some function of $\zetasym$ made to depend on the observed nonperturbative external state $|h \rangle$. The $d'(\zetasym,\left\{p_h\right\})$ is a modified definition of the fragmentation function.  The $X_h(\zetasym)$ could depend on any details of the nonperturbative state including its species, mass or, most relevantly for the current discussion, the number of hadrons in the state. 

The \emph{absolute} normalization of the hard part can be fixed by convention, but once a choice is made it must be kept fixed for all types of external states.  Factors should not move in and out of the hard part in a way that depends on the external nonperturbative states. 
If they do, then the hard part does not describe properties that are independent of the nonperturbative states.  
By arbitrarily choosing different $X_h(\zetasym)$'s for each type of nonperturbative final state, one may always force the fragmentation functions $d'(\zetasym,\left\{p_h\right\})$ to be any functions one desires. Then, the question of which definition is correct loses any meaning.

We believe that this is what the authors of Ref.~\cite{Pitonyak:2025lin} do to arrive at their \pitseven{\vers}. 
The manipulations there are expressed in the language of Jacobians, but the steps in going between Ref.~\cite[\pitseven{\vers}]{Pitonyak:2025lin} and Ref.~\cite[\pitnine{\vers}]{Pitonyak:2025lin} amount to redefining the fragmentation function within the factorization formula by shuffling factors back and forth between the fragmentation function and the hard part as in \eref{fact_formulapf} above. 
Moreover, the authors of Ref.~\cite{Pitonyak:2025lin} claim to have written, with their \pitseven{\vers}, a factorized formula for the cross section employing the standard parton model expression for the hard part.
But that is not the case. 
The extra hadron-number-dependent powers of $\zetasym$ have just been grouped 
into $\diff{\hat{z}}
{}/\hat{z}^{n-1}$ with the factors outside the parentheses.   
However, there is only one fragmenting parton, hence the integration upon its variables must not depend on the number $n$ of hadrons in the final state. 
The crucial point, in other words, is that the information about the dependence on the number of hadrons in the observed nonperturbative final state has been moved outside of the fragmentation function, thereby spoiling factorization. The presence of these spurious factors cannot be avoided by grouping them with factors labeled as being part of the integration measure.

Finally, the authors of Ref.~\cite{Pitonyak:2025lin} mention that when $n=2$ the dihadron fragmentation needs to be matched onto two single-hadron fragmentation functions at large $M_h$. We just point out here that nothing prevents one from expanding the $d(\zetasym,\left\{p_h\right\})$ from~\cite[Eq.~(3)]{Rogers:2024nhb} in single-hadron collinear fragmentation functions at large $M_h$ and performing the matching consistently. There is no difficulty with consistently matching single parton fragmentation at small $M_h$ with a large $M_h$ treatment.

If the alternative definition from Ref.~\cite{Pitonyak:2023gjx} is used, the problems with the factorization discussed above extend also to the treatment of evolution kernels. Renormalization factors are independent of the identity of nonperturbative external states, and this is central to their partonic interpretation, but the modified definition in Ref.~\cite{Pitonyak:2023gjx} forces them to acquire external state dependence. The definition in the last set of parentheses of~\cite[Eq.~(7)]{Pitonyak:2025lin}, when treated as the fundamental definition, suffers from the problems outlined in
Ref.~\cite[Sec.~VIII \& App.~C]{Rogers:2024nhb}.


\section{Number Densities}
\label{s.interp}

To compare the standard definition with the one promoted in Refs.~\cite[Sec.~IV]{Pitonyak:2023gjx,Pitonyak:2025lin}, we retrace here the steps for writing down a quark fragmentation function.  One starts with the general number density operator for hadron $h$ made from completely generic asymptotic out-states, 
\begin{equation}
a_{p_h,\text{out}}^\dagger a_{p_h,\text{out}}^{} = \frac{\diff{\widehat{N}}}{\diff{Y}{}} = \SumInt_{X}  | p_h, X \rangle \langle p_h, X | = (2 \pi)^3 2 E_h \frac{\diff{\widehat{N}}}{\diff[3]{\vect{p}_h}} = (2 \pi)^3 2 \xi \frac{\diff{\widehat{N}}}{\diff{\xi} \diff[2]{\T{p}{h,p}}} \, . \label{e.differential}
\end{equation}
It only involves the hadron momentum $p_h$, it has no knowledge about the existence of partons, and it makes no reference to a specific process. 
The far right is expressed using the parton momentum fraction $\xi$, but that is only possible because we specify the components of $p_h$ in a parton frame ($\vect{p}_{h,p}$) where there is a fixed external $k_{p}^+$. The full phase space is Lorentz invariant and independent of coordinates. 

To construct a fragmentation function, one takes the $\diff{\widehat{N}}{}/\diff{Y}{}$ expectation value between quark states and sets that equal to a function identified with the fragmentation function, up to a quark state normalization,
\begin{equation}
\langle \vect{k} | \frac{\diff{\widehat{N}}{}}{\diff{Y}{}} | \vect{k}' \rangle = \langle \vect{k} |  \vect{k}' \rangle \mathcal{H} \, d(\xi,-\xi \T{k}{H},p_h) . \label{e.differential2}
\end{equation}
The left side of the equality
has two separate parts: {\bf i)} the hadron number density operator from \eref{differential}, which only depends on external hadron states and makes no reference to quarks; {\bf ii)} the quark states 
which bring in the dependence on a quark momentum.  
On the right side of \eref{differential2}, one isolates all the dependence on the external states inside a function labeled the fragmentation function $d(\xi,-\xi \T{k}{H},p_h)$. Having all the dependence on nonperturbative external states isolated in a universal factor
is central to what it means that $d$ is
a fragmentation function. The rest of the factors on the right side are any leftover parts that are \emph{independent} of the external nonperturbative states. That includes the 
distribution $\langle \vect{k} |  \vect{k}' \rangle$ and a possible normalization factor $\mathcal{H}$ that fixes the integrals over $\vect{k'}$ that determines the normalization convention for the light-cone quark states.  
Since $\mathcal{H}$ is unrelated to the external nonperturbative states, and is not inside the fragmentation function, it ends up being absorbed into the hard parts of whatever factorization formulas involve the fragmentation function. 
Its practical effect is to fix the 
overall normalizations of the hard parts.\footnote{The overall normalization is fixed not only at the parton model level but for any perturbative order.} 
Since it goes into the hard part, $\mathcal{H}$ must be completely independent of the states in \eref{differential} if factorization is to be preserved, and it must be completely universal. Normally, one automatically just uses the most natural normalization, 
\begin{equation}
\mathcal{H} = (2 \pi)^3 2 \xi  \, , \label{e.basicH}
\end{equation}
because then hard parts in factorization have exactly the normalizations of partonic cross sections. 
With this choice, the
$\mathcal{H}$ brings the factorization formula as close as possible to the intuitive parton model. 
Other values for $\mathcal{H}$ are acceptable in principle \emph{so long as it is kept totally universal and independent of the external states that make up the hadron number density operator.}   
Using \eref{basicH}, we can write
\begin{equation}
\langle \vect{k} | (2 \pi)^3 2 E_h \frac{\diff{\widehat{N}}}{\diff[3]{\vect{p}_h}} | \vect{k}' \rangle = \langle \vect{k} |  \vect{k}' \rangle  (2 \pi)^3 2 \xi \, d(\xi,-\xi \T{k}{H},p_h) \, . \label{e.differential3}
\end{equation} 
This is the normal way to build up the definition of a fragmentation function. 
Equation~\eqref{e.differential3} is equivalent, for example, to Ref.~\cite[Eq.(4.5)]{Collins:1981uw} in the parton frame.\footnote{Note that, similarly to \eref{differential} above, it is the number density operator $a_A^\dagger a_A$ that appears in the quark matrix element in Ref.~\cite[Eq.(4.5)]{Collins:1981uw}, and the factor $1/2 \, P^+ (2 \pi)^{d-1}$ on the right-hand side just gives the $(2 \pi)^3 2 \xi$ in \eref{differential3} when using parton frame variables.} 
It is also equivalent to Ref.~\cite[Eq.~(12.35)]{Collins:2011qcdbook}. 
The definition is very intuitive and natural: One just takes the quark expectation value of the standard hadron number density operator. 
Using \eref{differential} we can rewrite  
\eref{differential3} as
\begin{equation}
\langle \vect{k} |  \frac{\diff{\widehat{N}}}{\diff{\xi} \diff[2]{\T{p}{h,p}}} | \vect{k}' \rangle = \langle \vect{k} |  \vect{k}' \rangle \, d(\xi,-\xi \T{k}{H},p_h) \, . \label{e.differential4}
\end{equation} 

Nothing in the discussion between \eref{differential} and \eref{differential4} depends on the number density operator in \eref{differential} being just for \emph{single} hadrons. 
Therefore, generalizing to multiple observed hadrons simply amounts to including the possibility that $\diff{\widehat{N}}/\diff{Y}$ 
has more than one hadron. 
The $n$-hadron generalization of \eref{differential} is 
\begin{align}
 \frac{\diff{\widehat{N_n}}}{\diff{Y}{}} &= a_{p_{h_1},\text{out}}^\dagger a_{p_{h_2},\text{out}}^\dagger \cdots a_{p_{h_n},\text{out}}^\dagger a_{p_{h_n},\text{out}}^{} \cdots a_{p_{h_2},\text{out}}^{} a_{p_{h_1},\text{out}}^{} \no
 &= \SumInt_{X}  | p_{h_1}, \dots, p_{h_n} X \rangle \langle p_{h_1}, \dots, p_{h_n}, X | = (2 \pi)^3 2 E_{h_1} \times \cdots \times (2 \pi)^3 2 E_{h_n} \frac{\diff{\widehat{N}_n}}{\diff[3]{\vect{p}_{h_1}} \cdots \diff[3]{\vect{p}_{h_n}}} \, . \label{e.differential5}
\end{align}
It is important that this operator is differential in multiple \emph{hadron} momenta, not parton momenta.
There are no other changes at all in the definition of the fragmentation function.  In particular, nothing on the right side of \eref{differential2} changes. All dependence on  nonperturbative external states remains isolated within the fragmentation function $d(\xi,-\xi \T{k}{H},\{p_h\})$. (See also Eq.(2.14) of Ref.~\cite{Collins:1994ax}.)

The number density interpretation here is also very direct and intuitive. The ordinary fragmentation function  applies in exactly the same familiar way, even if the observed final state is made of $n$ hadrons.
All aspects related to the partonic end of the fragmentation function are unchanged.  
This is the basic form that has been used most in past applications and is what we used in ~\cite[Sec.~V]{Rogers:2024nhb}.

By comparison, to recover the definition promoted in \cite[\pitfourteen{\vers}]{Pitonyak:2025lin}, one is forced to replace the standard quark normalization factor $\mathcal{H}$ in \eref{differential2} by the external-state-dependent factor
\begin{equation}
\mathcal{H} \to 
(2 \pi)^3 2 \xi_{1} \times \cdots \times (2 \pi)^3 2 \xi_{n} \propto (\xi)^n \, .
\end{equation}
However, this has a nontrivial dependence on the identity of the external nonperturbative state 
through the dependence on the number of hadrons $n$ in the state. The dependence on the external state is no longer isolated inside of $d(\xi,-\xi \T{k}{H},\{p_h\})$. This dependence gets placed inside the hard part when used in a factorization formula, and so it breaks factorization. 

We disagree with Ref.~\cite{Pitonyak:2025lin} that their \pitfourteen{\vers}, which is written as a density in $n$ separate \emph{parton} momenta, is standard or expected. While there are $n$ final state hadrons, there is only one fragmenting quark, so the appearance of $n$ copies of differentials in parton momentum fractions 
is very unexpected. In reality, in the standard definition the dependence on external \emph{hadron} momenta enters through the hadron number operator in \eref{differential5}, which makes no reference to partons. To obtain the definition for an $n$-hadron system,  
one does not first divide out the $(2 \pi)^3 2 \xi$ factor on both sides of \eref{differential3} for the special case of the single hadron and then simply append extra partonic differentials $\diff{\xi_i} \diff[2]{\T{p}{h_i,p}}$ to the left side to extend it to the case of multiple hadrons, but this appears to be what has happened with Ref.~\cite[\pitfourteen{\vers}]{Pitonyak:2025lin}.  

To try to better understand how to interpret the operator in Ref.~\cite[\pitfourteen{\vers}]{Pitonyak:2025lin}, we 
rewrite it using \eref{differential5} and
the parton frame relation $\xi_i k_p^+ = p_{h_i}^+$, namely 
\begin{align}
& \frac{\diff{\widehat{N}_n}}{\diff{\xi_1} \cdots \diff{\xi_n} \diff[2]{\T{p}{h_1,p}} \cdots \diff[2]{\T{p}{h_n,p}}} \sim \frac{1}{\xi^n}  \frac{\diff{\widehat{N}_n}}{\diff{Y}{}} \, ,  \label{e.pcmpsop}
\end{align}
where the ``$\sim$'' means that we do not write simple numerical factors that do not depend on $\xi$. Because it incorporates information about a \emph{parton} through the $1/\xi^n$, the operator in \eref{pcmpsop} is not a regular hadron number density operator like \eref{differential5}. Nor is it related to \eref{differential5} through a universal and state-independent normalization factor, like the $\mathcal{H}$ in \eref{basicH}.
Instead, it is related to $\diff{\widehat{N}_n}/\diff{Y}{}$ by a $1/\xi^n$ that intertwines external-state dependence (through $n$) with the partonic dependence on $\xi$. We have not found any natural way to interpret the use of such an operator within the context of single-parton fragmentation functions. 

The appearance of such factors is a consequence of attempting to force the single \emph{parton} momentum to play the role of the $n$ external \emph{hadron} momenta in the hadron number density operator, \eref{differential5}. 
Ultimately, that leads to the external-state-dependent factors in the hard part of Ref.~\cite[\pitseven{\vers}]{Pitonyak:2025lin} (or the extra $\xi$ in our \cite[Eq.~(140)]{Rogers:2024nhb}), which spoils the standard factorization of the cross section formula by making hard parts dependent on the identity of nonperturbative external states. 

\section{Conclusion}
\label{s.conc}
 
Moving forward, it is important to have clarity about which aspects of factorization, operator definitions, \textit{etc} are reasonably well established and which might need to be changed or modified. Our concern regarding Ref.~\cite{Pitonyak:2023gjx} is the likelihood that the sum rule there, due to its depiction as rigorous and indispensable for interpretations, will get incorporated into future constraints on theoretical treatments of fragmentation functions in studies of nonperturbative QCD. 
Nothing guarantees that Ref.~\cite[Eq.~(6)]{Pitonyak:2023gjx} relates fragmentation functions to actual hadron multiplicities, however, even to a rough approximation. Thus, 
using it to constrain models or nonperturbative treatments in this way would be problematic.
Meanwhile, the standard fragmentation function definition that has long been used in the past for nonperturbative modeling, for example in  Ref.~\cite[Eq.~(11)]{Matevosyan:2017alv}, is valid both technically and in the sense of interpretations. 
Past efforts like these do not need to be discarded on the basis of the claims made in Refs.~\cite{Pitonyak:2023gjx,Pitonyak:2025lin}. 

Such considerations become increasingly important as prospects for addressing fragmentation from first principles in nonperturbative QCD grow more feasible (see, e.g., the recent work in  Ref.~\cite{Galvez-Viruet:2025ket,Galvez-Viruet:2025rmy,Qian:2024gph,Grieninger:2024cdl,Grieninger:2024axp,Araz:2024bgg,vonKuk:2025kbv,Lee:2025okn,Kang:2025zto}). The independence of process-specific hard parts from universal, external-state-dependent fragmentation functions is central to the partonic interpretations that inspire this work.

\vskip 0.3in
\acknowledgments
T. Rogers was supported by the U.S. Department of Energy, Office of Science, Office of Nuclear Physics, under Award Number DE-SC0024715. 
A. Courtoy was supported  by the UNAM Grant No. DGAPA-PAPIIT  IN102225. This article was written without the use of AI.  

\bibliography{bibliography}

\end{document}